\documentclass[%
 reprint,
superscriptaddress,
 amsmath,amssymb,
 aps,
 prl,
floatfix,
]{revtex4-2}
\usepackage{cancel}
\usepackage[caption=false]{subfig}
\usepackage{graphicx}
\usepackage{dcolumn}
\usepackage{bm}
\usepackage[hidelinks]{hyperref}
\usepackage{xcolor}
\usepackage[normalem]{ulem}
\definecolor{ra}{rgb}{0.8, 0.0, 0.0}

\begin{document}

\preprint{APS/123-QED}

\title{Disordered Lattice Glass $\phi^{4}$ Quantum Field Theory}
\author{Dimitrios Bachtis}
\email{dimitrios.bachtis@phys.ens.fr}
\affiliation{Laboratoire de Physique de l'Ecole Normale Sup\'erieure, ENS, Universit\'e PSL,
CNRS, Sorbonne Universit\'e, Universit\'e de Paris, F-75005 Paris, France}

\include{ms.bib}

\date{July 09, 2024}

\begin{abstract}

We study numerically the three-dimensional $\phi^{4}$ spin glass, a prototypical disordered and discretized Euclidean field theory that manifests inhomogeneities in space and time but considers a homogeneous squared mass and lambda terms. The $\phi^{4}$ lattice glass field theory is a conceptual generalization of spin glasses to continuous degrees of freedom and we discuss the existence of a limit under which it formally reduces to the Edwards-Anderson model.  By defining four variants of an order parameter which are suitable for continuous spin glasses, we verify numerically the emergence of an overlap in absence of the magnetization thus confirming the presence of a spin glass phase transition for a value of a critical squared mass. We conclude by discussing how the $\phi^{4}$ spin glass can be utilized to address assumptions of complete homogeneity in space or time and how, in parallel to statistical physics, provides a suitable framework to investigate the nonperturbative dynamics of machine learning algorithms from the perspective of disordered lattice and constructive field theory.

\end{abstract}

\maketitle

\paragraph*{\label{sec:level1}Introduction.---}

Spin glasses~\citep{Mezard1987,RevModPhys.58.801}, such as the Edwards-Anderson~\citep{SFEdwards_1975,SFEdwards_1976} and Sherrington-Kirkpatrick~\citep{PhysRevLett.35.1792,PhysRevB.17.4384} models, have significantly influenced multiple aspects of the mathematical and physical sciences. In fact, the scientific impact of these systems reaches far beyond the topic of their initial conception. Spin glasses, which comprise a set of inhomogeneous interactions were originally proposed to describe a class of dilute magnetic alloys. It was soon realized that the same type of Hamiltonian present in the spin glass, is also suitable for the description of various research problems, which range from the dynamics of neural networks, to combinatorial optimization, and to the representation of financial markets. 

The fundamental characteristic behind the success of spin glasses is the presence of inhomogeneity in the system. Inhomogeneity  can be often interpreted as a coexistence of ferro- and anti-ferromagnetism that gives rise to a set of competing interactions which characterize the dynamics of a spin glass. Of notable mention is the fact that the concept of inhomogeneity is experimentally relevant: one imagines that real materials have some form of impurity or defect. It is therefore natural to pose the question of whether generalizations of the Edwards-Anderson or Sherrington-Kirkpatrick models should be studied, in the anticipation that they might potentially extend the already established success of spin glasses to a different class of scientific problems.  

In this manuscript, we study numerically the three-dimensional $\phi^{4}$ lattice glass field theory, a prototypical disordered and discretized Euclidean field theory that manifests inhomogeneities in space and time, but considers a homogeneous squared mass and lambda terms. To our knowledge, the $\phi^{4}$ lattice glass field theory has never been studied numerically. However, there exist proposals of soft-spin models within an analytical context to conduct mean-field studies of spin glasses. For instance, see Sompolinsky-Zippelius~\citep{PhysRevB.25.6860} and De Dominicis~\citep{PhysRevB.18.4913}. Such mean-field models generalize the Sherrington-Kirkpatrick model to quantum field theory and are to be considered conceptually equivalent to mean-field $\phi^{4}$ spin glasses.

\begin{figure}[t]
\includegraphics[width=8cm]{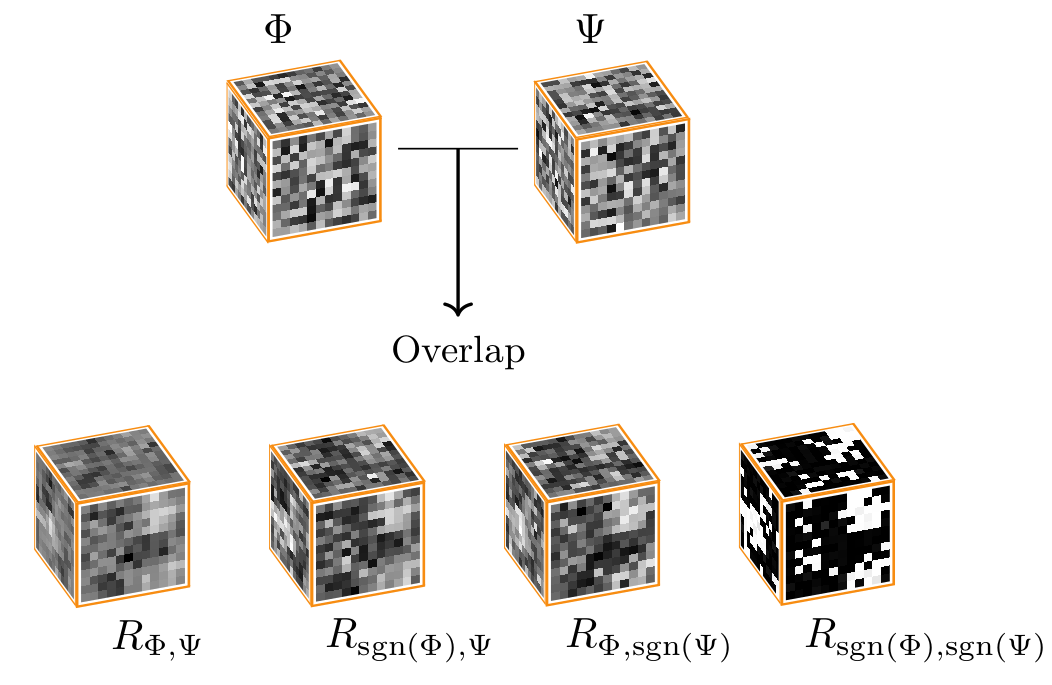}
\caption{\label{fig:fig1} The mapping of two replicas $\Phi$ and $\Psi$ of the $\phi^{4}$ lattice glass field theory into a set of overlap configurations $R$ using the four variants of overlap order parameters defined in the current manuscript.  }
\end{figure}

To solidify the connection between the $\phi^{4}$ glass studied here and short-range spin glasses we discuss the existence of a limit under which the $\phi^{4}$ lattice glass field theory, a system with continuous degrees of freedom, formally reduces to the Edwards-Anderson model.  We explore if we can define four variants of an overlap order parameter, suitable for the study of continuous spin glasses, and if we can verify numerically the presence of a spin glass phase for a value of a critical squared mass. Besides encoding Edwards-Anderson dynamics, the $\phi^{4}$ spin glass includes a crossover to a disordered quadratic action, and therefore provides a rich phenomenology to study numerically the spin glass phase transition~\citep{PhysRevLett.54.924}.

We conclude by discussing how extensions of concepts from spin glasses to lattice field theory are relevant for a diverse set of research problems. For instance, one can investigate the physical behavior of a lattice field theory which manifests inhomogeneity exclusively along certain dimensions of space and time.  Disordered quantum field theories are evidently a suitable framework for the description of various disordered materials.

The theory of disordered systems has been additionally successful in providing a theoretical framework to address research problems within computer science~\citep{10.5555/1592967}, such as those that emerge in the description of machine learning algorithms.  This class of problems usually considers datasets with continuous values in practical applications.  The $\phi^{4}$ spin glass preserves the structure of the Edwards-Anderson model and can be straightforwardly applied to model continuous datasets.  We specifically elaborate on how the $\phi^{4}$ lattice glass field theory can serve as a prototypical system to investigate the training dynamics of $\phi^{4}$ neural networks~\citep{PhysRevD.103.074510,Bachtis_2022}, a set of quantum field-theoretic machine learning algorithms which generalize restricted Boltzmann machines. We then discuss how $\phi^{4}$ spin glasses and $\phi^{4}$ neural networks, a class of Euclidean Markov random fields, can be potentially allocated further mathematical substance within the Nelson perspective of constructive quantum field theory~\citep{NELSON197397}. 

\paragraph*{\label{sec:level2}The $\phi^{4}$ lattice glass field theory.---} We define the three-dimensional $\phi^{4}$ lattice glass field theory with the lattice action $S$:
\begin{equation}\label{eq:oldaction}
S_{\Phi}=-\sum_{\langle ij \rangle} J_{ij} \phi_{i} \phi_{j} + \bigg(d+\frac{\mu^{2}}{2}\bigg) \sum_{i} \phi_{i}^{2}+\frac{\lambda}{4} \sum_{i} \phi_{i}^{4},
\end{equation}
where $d=3$ is the dimension of the system, $\mu^{2}$ and $\lambda$ are the squared mass and lambda couplings, and $\langle ij \rangle$ defines a nearest-neighbor interaction on a square lattice. The inhomogeneous couplings $J_{ij}$ are sampled uniformly as $J_{ij}=\pm 1$ and the set $\lbrace J_{ij}\rbrace$ defines a given realization of disorder. Any notion of inverse temperature is considered absorbed within the couplings. We denote a configuration as $\Phi$ and the value of a field at lattice site $i$ as $\phi_{i}$. The $\phi^{4}$ lattice glass field theory is a system with continuous degrees of freedom, namely $-\infty<\phi_{i}<\infty$. We simulate the system on a square lattice for a given lattice volume $V=L^{3}$, where  $L$ is the lattice size in each dimension. 

We remark that systems which manifest spin glass dynamics necessitate the simulation of multiple replicas to study their phase transition. We therefore consider two replicas of the system $\Phi$ and $\Psi$, with the same realization of disorder $\lbrace J_{ij}\rbrace$, and degrees of freedom $\phi_{i}$ and $\psi_{i}$, which define the joint Boltzmann probability distribution:
\begin{equation}\label{eq:origprob}
p_{\Phi_{i},\Psi_{j}}= \frac{\exp[-(S_{\Phi_{i}}+ S_{\Psi_{j}})]}{Z_{\Phi} Z_{\Psi}},
\end{equation}
where $Z_{\Phi}=\int_{-\infty}^{\infty} \exp[-S_{\Phi}] \prod_{i} d\phi_{i}$. We observe that $Z_{\Phi}Z_{\Psi}=Z_{\Phi}^{2}=Z_{\Psi}^{2} \equiv Z^{2}$. The replicas are not allowed to interact. It is straightforward to demonstrate, via the Hammersley-Clifford theorem, that the $\phi^{4}$ lattice glass field theory satisfies local, global, and pairwise Markov properties and that it is a Markov random field~\citep{PhysRevD.103.074510,Bachtis_2022}. This insight has been utilized to construct $\phi^{4}$ neural networks.

Due to the presence of disorder,  the expectation value of an arbitrary observable $O$ under the probability distribution $p_{\Phi}$ is calculated as:
\begin{equation}
[\langle O \rangle]_{J_{ij}}=\bigg[ \sum_{\Phi} p_{\Phi} O_{\Phi}\bigg]_{\lbrace J_{ij} \rbrace},
\end{equation}
where $\langle \rangle$ is a thermal average and $[]_{\lbrace J_{ij} \rbrace}$ denotes an averaging over the realizations of disorder $\lbrace J_{ij} \rbrace$ which are sampled probabilistically as discussed above. 

\paragraph*{\label{sec:level4}$\phi^{4}$ glass and the Edwards-Anderson model.---}It is useful to illustrate that the $\phi^{4}$ glass with $J_{ij}=\pm1$ considered here is a generalization of the Edwards-Anderson model  by demonstrating, in the limit $\lambda \rightarrow \infty$ and $\mu^{2} \rightarrow -\infty$, that it formally reduces to the Edwards-Anderson model. We recast the lattice action as:

\begin{align}\label{eq:newaction}
S_{\Phi} &= \frac{1}{2} \sum_{\langle ij \rangle, J_{ij}=1} (\phi_{i}-\phi_{j})^{2} \\ &+ \frac{1}{2}\sum_{\langle ij \rangle, J_{ij}=-1} (\phi_{i}+\phi_{j})^{2}  +\frac{\mu^{2}}{2} \sum_{i} \phi_{i}^{2}+\frac{\lambda}{4} \sum_{i} \phi_{i}^{4}.
\end{align} 

We are now able to set the following bound in the partition function:
\begin{equation}
Z_{\Phi} < \Bigg[\int_{-\infty}^{\infty}  \exp\bigg[-\bigg(\frac{\mu^{2}}{2}\phi^{2}_{i}+\frac{\lambda}{4}\phi^{4}_{i}\bigg)\bigg] d\phi_{i} \Bigg]^{V} < \infty,
\end{equation}
where the integral exists for $\lambda>0$.

Without loss of generality we set $\lambda/4 \rightarrow \lambda$ and $\mu^{2}/2=-2\lambda$ to obtain $(\mu^{2}/2)\phi^{2}+(\lambda/4)\phi^{4}\rightarrow\lambda(\phi^{2}-1)^{2}-\lambda$. We have now expressed the problem of taking the limits $\lambda\rightarrow \infty$  and $\mu^{2}\rightarrow-\infty$ in the original lattice action of Eq.~\ref{eq:oldaction} as the equivalent problem of taking the limit of only $\lambda \rightarrow \infty $.  We recall that 
\begin{equation}
\lim_{\lambda \rightarrow \infty} \frac{\sqrt{\lambda}}{\sqrt{\pi}} \exp[-\lambda(\phi^{2}-1)^{2}]= \delta(\phi^{2}-1),
\end{equation}

In the calculation of expectations of observables one obtains a measure where the only possible values for $\delta(\phi^{2}-1)$ are $\phi_{i}=\pm 1 \equiv s_{i}$, with $s$ denoting a binary spin. By substituting to the original lattice action we obtain $S=-\sum_{\langle ij \rangle} J_{ij} s_{i} s_{j}+c$, where $c$ denotes a constant value. The constant value cancels in the calculation of observables and we therefore recover, in the limit $\lambda \rightarrow \infty$ and $\mu^{2} \rightarrow -\infty$, the Edwards-Anderson model. Consequently, we expect the presence of a spin glass phase transition for the case of the $\phi^{4}$ lattice glass field theory, exactly in the same manner that we expect an Ising ferromagnetic transition for the homogeneous $\phi^{4}$ theory~\citep{PhysRevD.79.056008,arxiv.2205.08156}.
\begin{figure}[t]
\includegraphics[width=8.2cm]{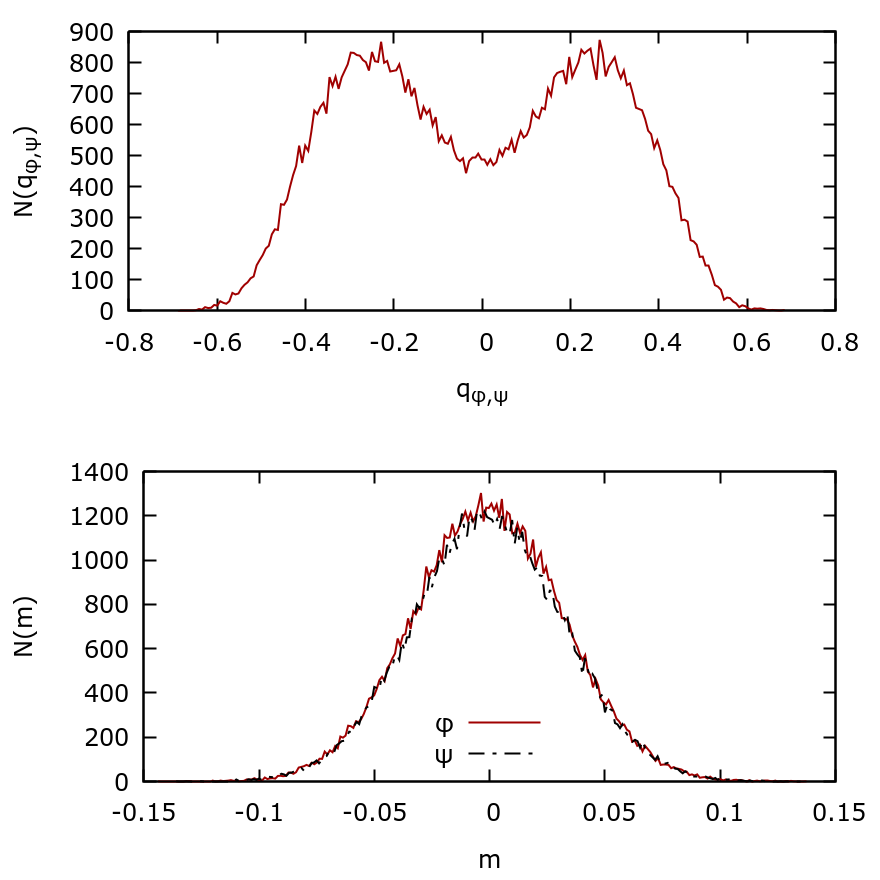}
\caption{\label{fig:fig2} Histograms $N(q_{\Phi,\Psi})$ of the overlap order parameter $q_{\Phi,\Psi}$ (top) and histograms $N(m)$ for the magnetization $m$ of each replica $\Phi$ or $\Psi$ (bottom). The results are obtained for a fixed realization of disorder at the value of the squared mass $\mu^{2}=-3.4$.}
\end{figure}

The order parameter of spin glasses is provided by the overlap function $q$~\cite{SFEdwards_1975,SFEdwards_1976,PhysRevLett.43.1754,PhysRevLett.50.1946,PhysRevLett.52.1156}, which is a measure of the similarity between two replicas. The overlap function is calculated based on a set of overlap degrees of freedom $r$, for example as $r_{i}=\phi_{i}\psi_{i}$. We remark that the set of $r_{i}=\phi_{i}\psi_{i}$ define an overlap configuration $R$. The consideration of the overlap function is crucial for the study of the system, because the correlation length $\xi$ can be directly observed on the overlap degrees of freedom $r$.  Since the $\phi^{4}$ lattice glass field theory is a system with continuous degrees of freedom $-\infty<\phi_{i}<\infty$, multiple overlap order parameters can be defined: it is not necessary for the considered order parameters to satisfy $q \in [0,1]$. In this manuscript we consider four variants:

\begin{gather}
q_{\Phi,\Psi}= \frac{1}{V} \sum_{i} \phi_{i} \psi_{i},\\
q_{\Phi,\textrm{sgn}(\Psi)}= \frac{1}{V} \sum_{i} \phi_{i} \textrm{sgn}(\psi_{i}),\\
q_{\textrm{sgn}(\Phi),\Psi}= \frac{1}{V} \sum_{i} \textrm{sgn}(\phi_{i}) \psi_{i},\\
q_{\textrm{sgn}(\Phi),\textrm{sgn}(\Psi)}= \frac{1}{V} \sum_{i} \textrm{sgn}(\phi_{i}) \textrm{sgn}(\psi_{i}),
\end{gather}
where $\textrm{sgn}$ denotes the sign function. The first order parameter $q_{\Phi,\Psi}$ is a straightforward definition for continuous degrees of freedom with the characteristic that it does not consider any sophisticated rescaling for the magnitude of the overlap degree of freedom $r_{i}=\phi_{i}\psi_{i}$. The second $q_{\Phi,\textrm{sgn}(\Psi)}$ and third $q_{\textrm{sgn}(\Phi),\Psi}$ definitions have the characteristic that they are able to encode the correlation length on an original configuration using the original magnitude of the fields, either of the $\Psi$ or $\Phi$ replica. The last definition $q_{\textrm{sgn}(\Phi),\textrm{sgn}(\Psi)}$, which coincides with the overlap order parameter of the Edwards-Anderson model, is mathematically unjustified because the $\phi^{4}$ glass is a system with continuous degrees of freedom and one must take into consideration the magnitude of the fields.  Nevertheless, we will demonstrate that all four variants of the overlap order parameter, see Fig.~{\ref{fig:fig1}}, act as phase indicators for the $\phi^{4}$ spin glass transition.

\begin{figure}[t]
\includegraphics[width=8.2cm]{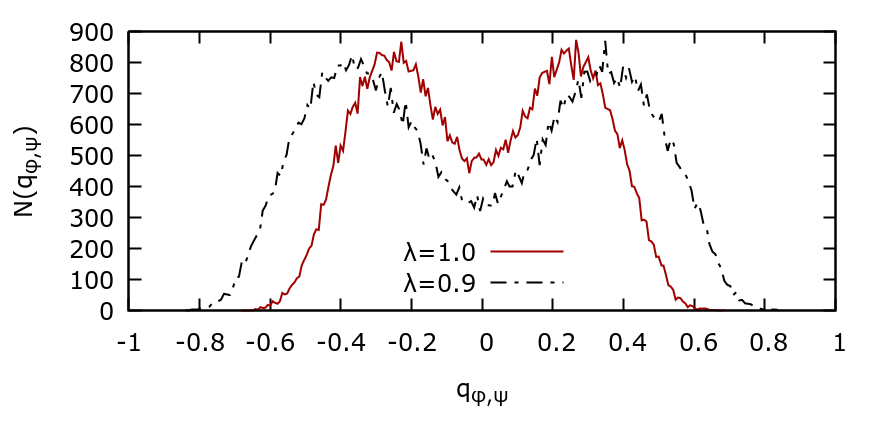}
\caption{\label{fig:fig3}Histograms $N(q_{\Phi,\Psi})$ of the overlap order parameter $q_{\Phi,\Psi}$ for a fixed realization of disorder and $\mu^{2}=-3.4$, but for different values of lambda, namely $\lambda=1.0$ and $\lambda=0.9$.}
\end{figure}

\paragraph*{\label{sec:level3}The $\phi^{4}$ spin glass phase transition.---} We study numerically the $\phi^{4}$ spin glass phase transition with the implementation of replica exchange Monte Carlo methods~\citep{PhysRevLett.57.2607,Marinari_1992,doi:10.1143/JPSJ.65.1604}. Specifically, we consider $N=40$ simulations for each realization of disorder $\lbrace J_{ij} \rbrace$. We clarify that a smaller number of simulations can be used to study the phase transition of the  system. Nevertheless, since this manuscript conducts, to our knowledge, the first simulation of the $\phi^{4}$ spin glass we consider it important to sample a large region of parameter space. The simulations are conducted for a fixed value of $\lambda=1.0$ and for $N$ values of the squared mass $\mu^{2}$ in the range $\mu^{2} \in [-3.8,-1.85]$. We remark that the $\phi^{4}$ spin glass is more difficult to study numerically than discrete spin glasses, such as the Edwards-Anderson model, due to the fact that one needs to be additionally updating with Monte Carlo simulations the magnitude of the fields. We therefore consider lattices for volumes $V=12^{3},10^{3},8^{3},6^{3}$, and $10$ realizations of disorder. For replica exchange Monte Carlo methods to be successful we impose the constraint that the consecutive simulations for adjacent values of the squared mass $\mu^{2}$ have overlapping histograms of the lattice action.

\begin{figure}[t]
\includegraphics[width=8.2cm]{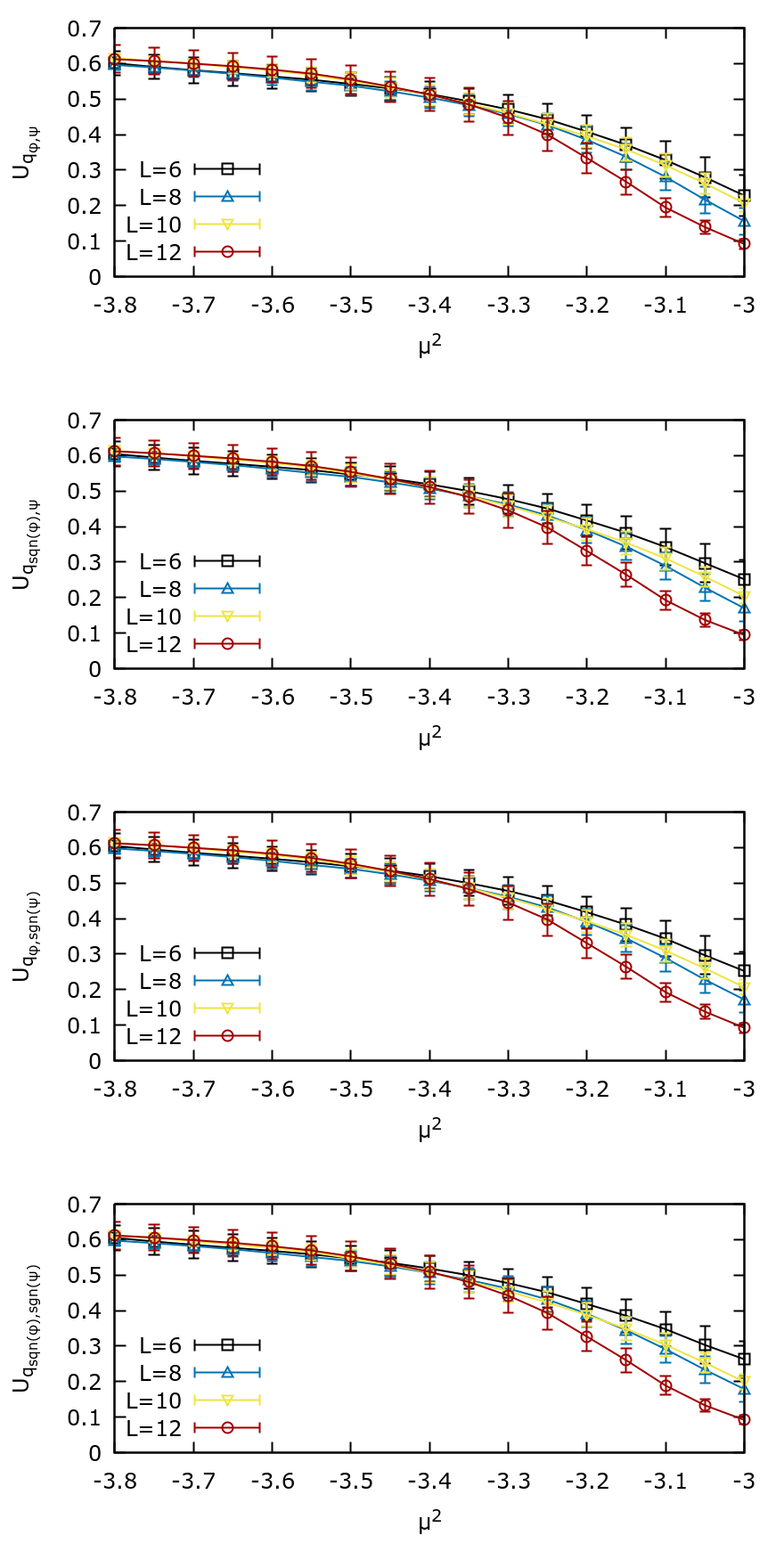}
\caption{\label{fig:binder} The Binder cumulant $U_{q}$ as calculated for the four variants of the overlap order parameter $q$ versus the value of the squared mass $\mu^{2}$. }
\end{figure}

In order to establish the presence of a $\phi^{4}$ spin glass phase transition in absence of any magnetization we consider a fixed realization of disorder $\lbrace J_{ij} \rbrace$ for $V=8^{3}$ and we calculate the histograms of the overlap order parameter $q_{\Phi,\Psi}$ in addition to the histograms of the magnetizations $m_{\Phi}$, $m_{\Psi}$ for each replica $\Phi$ and $\Psi$, where $m_{\Phi}=(1/V)\sum_{i}\phi_{i}$. The results, which consider the value $\mu^{2}=-3.4$ for the squared mass, are depicted in Fig.~\ref{fig:fig2}. We observe the emergence of an overlap order parameter in absence of the magnetization, thus verifying the analytically predicted presence of a $\phi^{4}$ spin glass phase transition for the system. The $\phi^{4}$ spin glass does not undergo a ferromagnetic phase transition in the space of the original lattice action due to the spontaneous breaking of the $Z_{2}$ symmetry. Nevertheless, one can consider that a $Z_{2}$ symmetry breaking phase transition is observed on the space of the overlap configurations, which define an effective system, see Refs.~\citep{PhysRevLett.55.2606,PhysRevLett.112.175701,PhysRevB.98.174205,PhysRevB.98.174206,Bachtis_2024over,PhysRevB.37.7745,bachtis2023inv}.

In the context of the homogeneous $\phi^{4}$ theory~\citep{PhysRevD.79.056008,PhysRevLett.128.081603,arxiv.2205.08156}, a given value of $\lambda>0$ gives rise to a ferromagnetic phase transition for a unique value of $\mu^{2}_{c}<0$. The set of $\lambda$ and $\mu_{c}^{2}$ then define a critical line or curve for the system. It is of interest to conduct a proof-of-principle demonstration to verify the presence of such a critical line or curve also for the case of the $\phi^{4}$ spin glass. Namely, we aim to provide evidence that  a given value of $\lambda>0$ gives rise to a spin glass phase transition for a unique value of $\mu^{2}_{c}<0$. We consider the same fixed realization of disorder $\lbrace J_{ij} \rbrace$ as above, and the same value of the squared mass $\mu^{2}=-3.4$, but change the value of $\lambda=1.0$ to $\lambda=0.9$. We then calculate the histograms of the overlap order parameter $q_{\Phi,\Psi}$ for $\lambda=0.9$ and we compare them against the histograms of the overlap order parameter $q_{\Phi,\Psi}$ for $\lambda=1.0$. The results are depicted in Fig.~\ref{fig:fig3}. We observe, for the same realization of disorder, that the histograms  for $\lambda=0.9$ provide larger values of the overlap order parameter $q_{\Phi,\Psi}$, thus indicating that the spin glass phase transition has been shifted to a different value of a pseudocritical $\mu^{2}_{c}$. 

We remark that the above calculations are proof-of-principle demonstrations in order to understand qualitatively the physical behavior of the $\phi^{4}$ spin glass. To study systems which manifest spin glass dynamics it is a necessity to be averaging over the realizations of disorder in order to calculate quantities that are physically meaningful for the system. A quantity that can be employed to study spin glass phase transitions is the Binder cumulant~\citep{PhysRevLett.47.693} which is defined as:
\begin{equation}
U_{q}=1-\frac{[\langle q^{4} \rangle]_{\lbrace J_{ij}\rbrace}}{3[\langle q^{2} \rangle ]^{2}_{\lbrace J_{ij} \rbrace}}.
\end{equation}

We conduct the calculation of the Binder cumulant for all four variants of the overlap function $q_{\Phi,\Psi}$, 
$q_{\Phi,\textrm{sgn}(\Psi)}$, $q_{\textrm{sgn}(\Phi),\Psi}$, $q_{\textrm{sgn}(\Phi),\textrm{sgn}(\Psi)}$. The results are depicted in Fig.~\ref{fig:binder}. We observe the anticipated physical behavior for all definitions of the overlap function. We remark that statistical errors are affected by the number of realizations of disorder. The Binder cumulant indicates intersections for all four variants of the overlap order parameter which serve an estimate for the location of the critical region of the $\phi^{4}$ spin glass phase transition for $\lambda=1.0$ and the considered type of realization of disorder.  

We additionally observe that, for the same data, it is possible that the presence of a fixed point could slightly vary between the different definitions of the overlap order parameter. It is therefore of interest to explore how different definitions of the overlap order parameter for continuous spin glasses affect the location of the critical point, as well as the calculation of the critical exponents. Such studies require extensive simulations and are beyond the scope of this exploratory work. We emphasize that one should be cautious when calculating critical quantities as $\lambda \rightarrow 0^{+}$ due to the crossover to a disordered action with a quadratic term which is expected to affect calculations of critical quantities.

\paragraph*{\label{sec:level5}Conclusions.---} We have studied the three-dimensional $\phi^{4}$ spin glass, a prototypical disordered and discretized Euclidean field theory that manifests inhomogeneities in space and time but considers a homogeneous squared mass and lambda terms. To our knowledge, the $\phi^{4}$ lattice glass field theory has never been studied numerically. Here, we employed Monte Carlo simulations to provide numerical evidence for the location of the critical region of the $\phi^{4}$ spin glass. We introduced four variants of an overlap order parameter which are suitable for the study of spin glasses with continuous degrees of freedom. We additionally discussed the existence of a limit under which the $\phi^{4}$ lattice glass field theory formally reduces to the Edwards-Anderson model. Besides encoding Edwards-Anderson dynamics on continuous degrees of freedom, the $\phi^{4}$ glass includes a crossover to a disordered quadratic action, and therefore provides a rich phenomenology to study numerically the spin glass phase transition.

The extension of numerical concepts from spin glasses to lattice quantum field theories opens up the opportunity to explore a set of diverse research problems. For instance, one can investigate the behavior of a lattice field theory when introducing inhomogeneity only along a certain dimension of spacetime. Disordered systems have additionally been successful in providing a theoretical framework to understand the dynamics which describe the learning of neural networks.  The $\phi^{4}$ spin glass can then be employed, as a prototypical system, to investigate the learning phase transitions~\citep{bachtis2024casc} during the training of $\phi^{4}$ neural networks~\citep{PhysRevD.103.074510,Bachtis_2022}. The $\phi^{4}$ spin glass and the $\phi^{4}$ neural networks additionally provide a set of algorithms that are capable of modelling continuous data but simultaneously preserve the structure of spin glasses, such as the Edwards-Anderson and Sherrington-Kirkpatrick models. A final research question, relating to the Nelson perspective of constructive quantum field theory~\citep{NELSON197397}, is whether $\phi^{4}$ glasses and $\phi^{4}$ neural networks, a class of Markov random fields, could ever be assigned any further mathematical substance by being rigorously constructed as quantum field theories in Minkowski space starting from their Markov random field interpretation in Euclidean space.

\paragraph*{\label{sec:level6}Acknowledgements.---}The author thanks Giulio Biroli for interesting discussions and for bringing to my attention literature pertinent to mean-field $\phi^{4}$ glasses. The author acknowledges support from the CFM-ENS Data Science Chair and PRAIRIE (the PaRis Artificial Intelligence Research InstitutE).
\newpage

\bibliography{ms}

\end{document}